\documentclass[runningheads]{llncs}
\usepackage{graphicx}
\usepackage{subfigure}
\usepackage{tabularx}
\usepackage{amsmath}
\usepackage{hyperref}
\hypersetup{
    colorlinks=true,
    linkcolor=blue,
    filecolor=blue,      
    urlcolor=blue,
    citecolor=cyan,
}
\begin{document}
\title{Nuclei Grading of Clear Cell Renal Cell Carcinoma in Histopathological Image by Composite High-Resolution Network}
% \author{Anonymous}
\author{Zeyu Gao\inst{1,2} \and
Jiangbo Shi\inst{1,2} \and
Xianli Zhang\inst{1,2} \and
Yang Li\inst{1,2} \and
Haichuan Zhang\inst{1,2} \and
Jialun Wu\inst{1,2} \and
Chunbao Wang\inst{3} \and
Deyu Meng\inst{2,4}
Chen Li\inst{1,2}}

\authorrunning{Z. Gao et al.}
% \authorrunning{Anonymous}

\institute{School of Computer Science and Technology, Xi'an Jiaotong University, Xi'an, Shaanxi 710049, China \and
National Engineering Lab for Big Data Analytics, Xi'an Jiaotong University, Xi'an, Shaanxi 710049, China
\and Department of Pathology, the First Affiliated Hospital of Xi’an Jiaotong University, Xi’an, 710061, China \and School of Mathematics and Statistics, Xi'an Jiaotong University, Xi'an, Shaanxi 710049, China\\
\email{gzy4119105156@stu.xjtu.edu.cn}}

% \institute{Anonymous Organization \\
% \email{***@*****.***}}

\titlerunning{Nuclei Grading of ccRCC in Histopathological Image by CHR-Net}
\maketitle              % typeset the header of the contribution
\begin{abstract}

The grade of clear cell renal cell carcinoma (ccRCC) is a critical prognostic factor, making ccRCC nuclei grading a crucial task in RCC pathology analysis.
Computer-aided nuclei grading aims to improve pathologists' work efficiency while reducing their misdiagnosis rate by automatically identifying the grades of tumor nuclei within histopathological images.
Such a task requires precisely segment and accurately classify the nuclei.
However, most of the existing nuclei segmentation and classification methods can not handle the inter-class similarity property of nuclei grading, thus can not be directly applied to the ccRCC grading task.
In this paper, we propose a Composite High-Resolution Network for ccRCC nuclei grading. 
Specifically, we propose a segmentation network called W-Net that can separate the clustered nuclei.
Then, we recast the fine-grained classification of nuclei to two cross-category classification tasks, based on two high-resolution feature extractors (HRFEs) which are proposed for learning these two tasks.
The two HRFEs share the same backbone encoder with W-Net by a composite connection so that meaningful features for the segmentation task can be inherited for the classification task.
Last, a head-fusion block is applied to generate the predicted label of each nucleus. 
Furthermore, we introduce a dataset for ccRCC nuclei grading, containing 1000 image patches with 70945 annotated nuclei. 
We demonstrate that our proposed method achieves state-of-the-art performance compared to existing methods on this large ccRCC grading dataset.

\end{abstract}
\section{Introduction}
\begin{figure}[h]
\includegraphics[width=\textwidth]{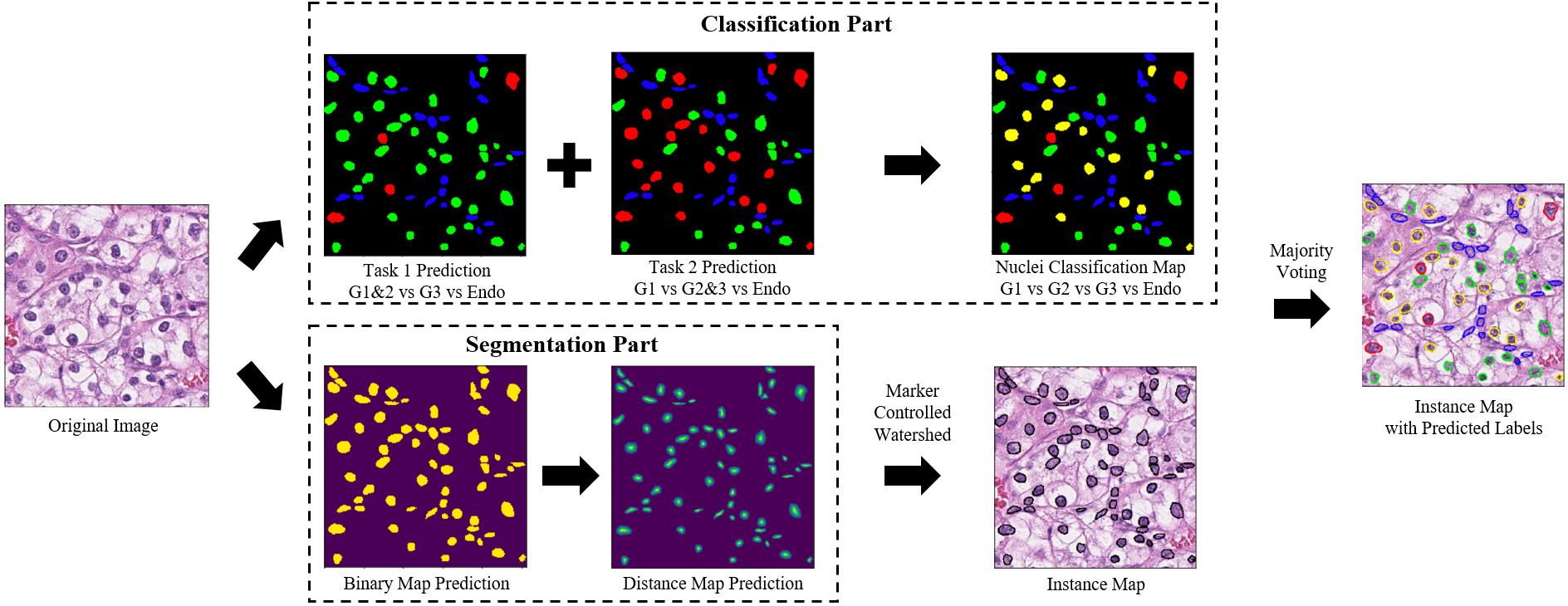}
\caption{Overview of the proposed nuclei grading method. G1, G2, and G3 denote tumor nuclei with grade 1 to 3 (green, yellow and red), Endo is the endothelial nuclei (blue).} \label{fig1}
\end{figure}
% What is cancer grading and why its important for ccRCC, introduce the isup grading system. The drawbacks of human grading system, and introduce how Precision medicine can help people.
Clear cell renal cell carcinoma (ccRCC) is the most common subtype of renal cell carcinoma (RCC), making up about 80\% of all cases. 
Grading of ccRCC has been recognized as a critical prognostic factor, where the overall five-year cancer-specific survival rate varies from 20\% to 90\% with different grades \cite{ccRCCGrading}.
Recently, a novel grading guideline only bases upon nucleoli prominence for grades 1–3 is recommended and validated for both ccRCC and papillary (p) RCC by the ISUP and the WHO, namely ISUP/WHO grading \cite{isup}.
Practically, it is unfeasible for pathologists to recognize every single nucleus considering the large size of a tumor slide.
Instead, pathologists usually select a few regions for diagnosis, which is rough and may missing some critical diagnostic information.
Moreover, pathologists cannot avoid subjectivity and randomness, which may result in inconsistent diagnoses.
To tackle these limitations, this paper focuses on developing an accurate computer-aided diagnosis (CAD) system for automatically identify the grade of each nucleus in ccRCC.

% The key points for nuclei grading
Accurate nuclei grading of ccRCC relies on the precise segmentation and the fine-grained classification of each nucleus.
For precise segmentation, one popular paradigm is to model the segmentation task as a distance regression task via various methods \cite{DIST,centerVectorEncoding,Hovernet}, which has shown superior performance in nuclei segmentation.
%, especially in separating torching nuclei.
However, directly learn the distance map from the original image is very challenging, which requires a model to focus on both the foreground-background dissimilarity and the spatial information (shape and distance) of each foreground object.
As for the nucleus classification, most advanced frameworks generally add a nuclei classification branch to a segmentation network \cite{Hovernet,isbi2019}.
Unlike the coarse-grained classification (\textit{e.g.,} tumor vs. other types), the fine-grained ccRCC nuclei classification is more challenging because of the inter-class similarity property.
Moreover, few existing datasets contain fine-grained labels for the nuclei grading task.

% Fix these limitation
To tackle all the aforementioned limitations, we propose a novel method named Composite High-Resolution Network (CHR-Net) for the ccRCC nuclei grading task, 
as well as introduce a new dataset that contains segmentation annotation and fine-grained labels of ccRCC nuclei.
The proposed CHR-Net consists of a backbone encoder to extract features from the input image, a distance regression-based instance segmentation branch to segment each nucleus precisely, and a fine-grained classification branch for predicting the grade of each nucleus.
The overall framework of the proposed method is shown in Fig. \ref{fig1}.

Specifically, to make the learning process of distance regression-based instance segmentation task easy and stable, we propose a two-stage network composes of two U-Net-like networks, namely W-Net. 
The first network is used to predict the binary maps (\textit{i.e.,} background vs. foreground) that contain the shape information of nuclei. 
The second is a lightweight U-net, which takes the binary maps as inputs and predicts the distance maps.
%As the binary maps contain the shape information of the nucleus, predicting the distance map from it is easier than that from the original image.

In the classification branch, considering the inter-class similarity property of ccRCC nuclei grading, we divide the fine-grained classification task into two sub-tasks based on the ISUP/WHO grading guideline first and then generate the final predictions by a simple fusion block.
Another key point for the fine-grained classification is how to maintain fine-grained features in the representation. 
To this end, we propose a high-resolution feature extractor (HRFE) based on HRNet \cite{HRNet} to maintains high-resolution representations throughout the whole branch of classification and a composite connection \cite{CBNet} to inherit meaningful features from the backbone encoder of the W-Net. 

As a part of this work, we introduce a new dataset for ccRCC nuclei grading, which contains 70945 exhaustively annotated nuclei within four nuclei types (grade 1-3, endothelial). Extensive experiments conducted on this dataset demonstrate the state-of-the-art performance of CHR-Net.

\begin{figure}[t]
\includegraphics[width=\textwidth]{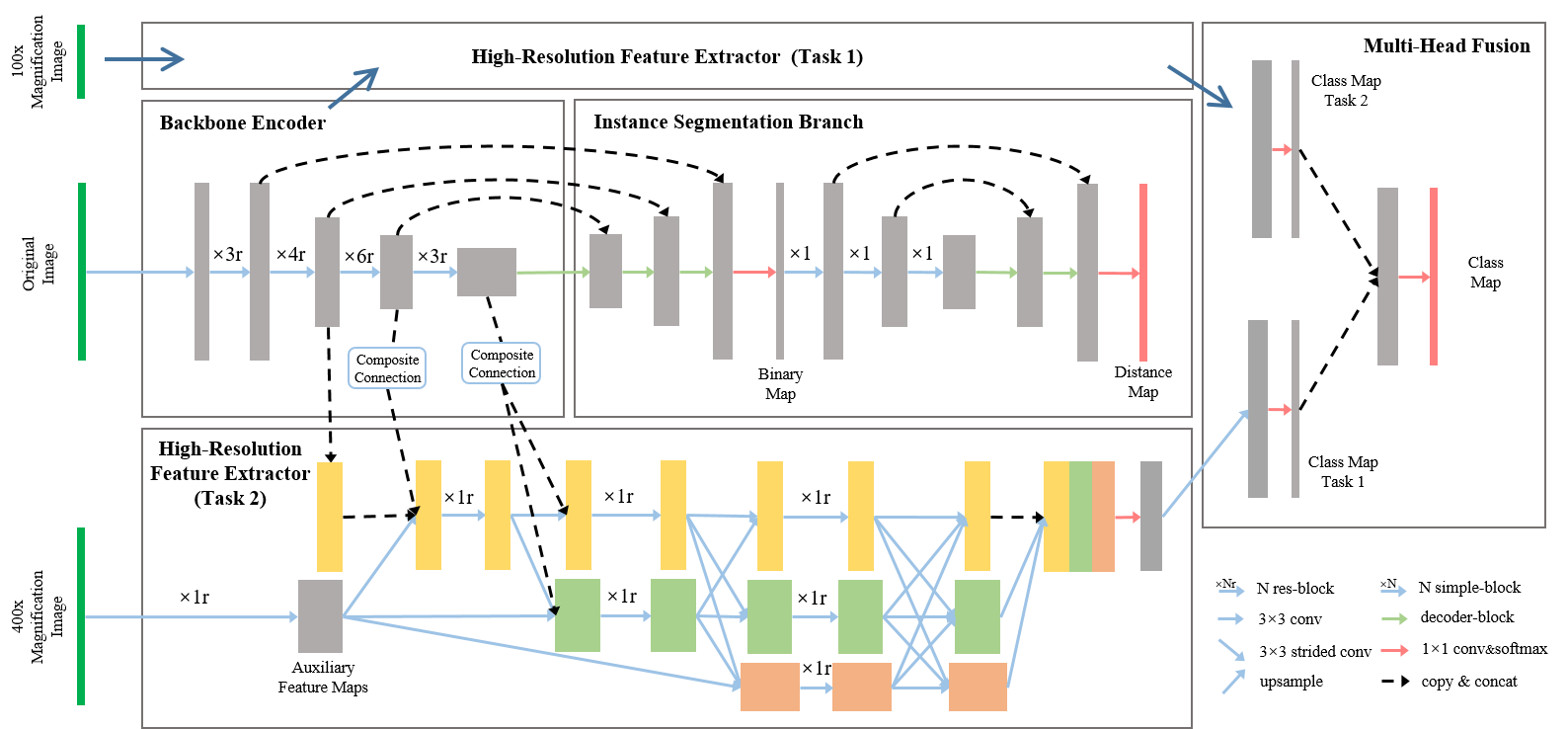}
\caption{Architecture details of CHR-Net. The High-Resolution Feature Extractor of Task 1 is hidden for brief display and has the same structure with Task 2.} \label{fig2}
\end{figure}

\section{Method}
In this section we expatiate the proposed CHR-Net, and the architecture details are shown in Fig. \ref{fig2}. 
The backbone encoder of W-Net is a ResNet-34 \cite{resnet} with global context attention \cite{GCNet}. 
The res-block used in HRFE is the same as the backbone encoder, which contains two $3\times3$ convolution layers and one shortcut layer. 
The simple-block has two $3\times3$ convolution layers, and the decoder-block contains one upsample, one shortcut, and two $3\times3$ convolution layers.

\subsection{Instance Segmentation with W-Net}

For instance segmentation, the distance regression schema has been proven to be effective. DIST \cite{DIST} predicts the distance between nuclei pixels and their nearest background pixel. 
Center Vector Encoding \cite{centerVectorEncoding} and HoVer-Net \cite{Hovernet} further considered the horizontal and vertical distances. 
These works directly learn the mapping between the original image and the distance map through an end-to-end neural network, which is hard to capture all valuable features simultaneously.

Instead, we utilize a two-stage learning framework W-Net to solve this complex task by dividing it into two sub-tasks. This insight is similar to curriculum learning \cite{curriculum_learning} which has been proven to be effective in histopathology image \cite{two_stage_learning}. In the first stage, a U-Net with ResNet-34 backbone encoder is utilized to mine the dissimilarity between the nuclei and the background while generating the binary maps. 
In the second stage, we apply a lightweight U-net (LU-Net) to extract the shape and distance-related features from the generated binary maps, meanwhile, convert the binary maps into the distance maps.
% predict the binary maps from the original images . 

% Recently, \cite{BendLoss} proposes to use the prior shape information from the original image to learn the distance 
% maps by giving high penalties to the contour points with high curvature. We assume, if the binary maps of the original image are well-learned by our model, it will be easier to capture the shape information from the binary maps than the original image. Therefore, 

% Then a lightweight U-Net (named as LU-Net) which only has three simple encoder blocks is used to convert the binary segmentation results into the regression-based distance maps.

\begin{figure}[t]
\centering
\subfigure[]{
\includegraphics[width=0.2\textwidth]{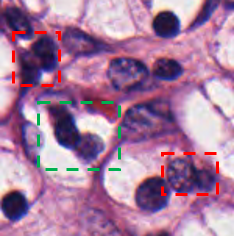}
}
\subfigure[]{
\includegraphics[width=0.2\textwidth]{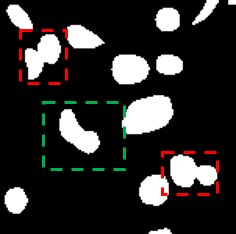}
}
\subfigure[]{
\includegraphics[width=0.2\textwidth]{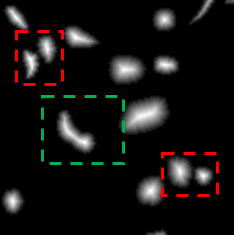}
}
\subfigure[]{
\includegraphics[width=0.2\textwidth]{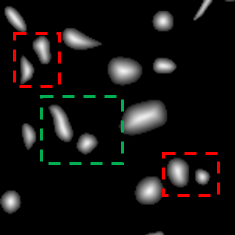}
}
\caption{Examples of touching nuclei. (a) Original image sub-patch. (b) Ground-truth binary map. (c) Ground-truth distance map. (d) Predicted distance map. Red rectangles indicate touching nuclei. The green rectangle represent a mislabeled nucleus which should be marked as two nuclei.} \label{fig3}
\end{figure}

As shown in Fig. \ref{fig3}(b), the shape of nuclei are mostly round or oval, but the touching nuclei have an extremely different pattern. We can see from Fig. \ref{fig3}(d) that by leveraging the strict mapping between the nuclei binary map and their distance map, the LU-Net bias the W-Net to separate the touching nuclei, even for the mislabeled one. Since the value of the distance map is continuous, L1 loss $L_{dist}$ is adopted for the regression task in the second stage. And for the binary classification in the first stage, the binary cross-entropy loss $L_{bc}$ is used.

To construct the distance map of each image, we adopt the same way as DIST \cite{DIST}. For the pixels of a nucleus, the corresponding values are defined by the distance between each pixel and the closest background pixel. Then we normalize the values of each nucleus to [0, 1].

\subsection{Composite High-Resolution Feature Extractor}
For nuclei classification, researchers follow the segmentation or detection then classification schema \cite{SP2016,2016lsDC} at the beginning. Recently, some researches raise that solve both tasks (i.e. segmentation and classification) in a unified framework can improve the performance of both tasks. Besides, some public dataset are collected for nuclei classification, such as CoNSeP \cite{Hovernet}, PanNuke \cite{gamper2020pannuke} and MoNuSAC \cite{monusac2020}. 
Despite some works that pay attention to nuclei classification, most of them are coarse-grained and rarely focus on the fine-grained nuclei classification that grading the tumor nuclei by their appearance.

Furthermore, a major property of nuclei grading is the inter-class similarity that also brings a challenge in nuclei classification, and it appears when observing the tumor nuclei at different magnifications. 
Specifically, the ISUP grading system for ccRCC is based on the visibility of the nucleoli. 
According to the grading principle, a tumor nuclei belongs to grade 1 if its nucleoli are invisible or small at 400x magnification, it belongs to grade 2 if it has conspicuous nucleoli at 400x magnification but inconspicuous at 100x, and belongs to grade 3 if its nucleoli are visible at 100x magnification. 
However, the grade 2 tumor nuclei is almost same as grade 1 at 100x magnification, as grade 3 at 400x magnification, thus leads inaccurate classification.
% However, the tumor nuclei in grade 2 and grade 3 or grade 1 and grade 2 may be almost the same at 400x and 100x magnification, respectively, thus leads inaccurate classification.
Therefore, ignoring this property may increase the uncertainty of the model, thereby reducing the model performance.

To address this issue, we design two cross-category classification tasks with different auxiliary inputs, and then aggregate the probability maps of these two tasks by a $1\times1$ convolution layer to generate the final classification output. 
The first task merges grade 1 and grade 2 into one category, and the second one merges grade 2 and grade 3. Meanwhile, the feature maps of 100x and 400x magnification images are taken as two auxiliary inputs, respectively. 
The original images are cropped from the 400x magnification WSIs, then we resize these images to 1/16 of the original size to obtain the 100x magnification images.
% The original images are cropped from the 40x objective magnification WSIs, then we resize these images to 1/16 of the original size to obtain the 10x objective magnification images. Note that, 400x magnification of the microscope indicates 10x eyepiece and 40x objective magnification, 100x magnification stands for 10x eyepiece and 10x objective magnification.

Another important point for fine-grained nuclei classification is to extract the semantically strong and spatially precise high-resolution representations. 
The U-net-like networks, which encode the input image with the high-to-low resolution convolutions and then recover the high-resolution representation from the low-resolution feature maps, have been successfully applied to nuclei segmentation.
However, due to the loss of spatial details, the low-resolution feature maps are insufficient for the nuclei classification, especially for small nuclei, even though applying the skip-connection in the networks. 
Thereby, we propose the HRFE, which consists of a parallel high-to-low resolution convolutions network with three multi-resolution streams for extracting high-resolution features. 
The structure of HRFE is illustrated in Fig. \ref{fig2}. Different from HRNet \cite{HRNet} that has only one input,  the original inputs of two HRFEs are 1) feature maps from the backbone encoder, and 2) the auxiliary feature maps from multi-scale images for different tasks. 
Furthermore, we adopt an accelerated version of the composite connection \cite{CBNet} that consists of an upsample and a $1 \times 1$ convolution layer with batch normalization to combine two HRFEs with the backbone encoder for the feature sharing of the multi-tasks.

% the composite connection \cite{CBNet} which consists of one upsample and one $1\times1$ convolution layer with batch normalization is used to boost the model performance by assembling multiple backbones

The loss function of the classification part is composed of three terms, the categorical cross-entropy losses of the first task $L_{mc1}$, the second task $L_{mc2}$ and the final classification output $L_{mcf}$. 
Finally, the overall loss function of CHR-Net is defined as:
\begin{equation}
    L = \underset{segmentation\, part}{\underbrace{\lambda _{bc}L_{bc} + \lambda _{dist}L_{dist}}} + \underset{classification\, part}{\underbrace{\lambda _{mc1}L_{mc1} + \lambda _{mc2}L_{mc2} + 
    \lambda _{mcf}L_{mcf}}},
\label{loss}
\end{equation}
where $\lambda _{bc}, ..., \lambda _{mcf}$ are parameters of each associated loss for weight control. Empirically, we set $\lambda _{dist} = 2$ and the others to 1.

\subsection{Post Processing}
With the prediction of the binary and distance map, we perform a marker-controlled watershed to generate the instance segmentation map of each test image. Particularly, the same strategy of \cite{DIST} is used to find all local maximums of the distance map, and then we take these local maximums as the marker to determine how to split the binary map.

To precisely identify the type of each nucleus, it is necessary to aggregate pixel-level type predictions to generate the type prediction of each nucleus instance. With the processed instance segmentation map, we use majority voting to obtain the most frequently appeared class of each nucleus instance in the nuclei classification map.

\section{Experiments and Results}

\subsection{Dataset}

A dataset for nuclei grading of ccRCC is proposed.
% This dataset is a modified version of \cite{datasetpaper}, with more accurate boundaries and balanced data distribution. 
To balance the data distribution of this dataset, we also include a part of papillary (p) RCC image patches, which follow the same nuclei grading system and have more high-grade nuclei.
The entire dataset consists of 1000 H\&E stained image patches with a resolution of $512\times512$. 
Two experienced pathologists are invited to select regions of interest (tumor regions) from 150 ccRCC, 50 pRCC WSIs derived from the KIRC, KIRP project of The Cancer Genome Atlas (TCGA) and scanned at 40x objective magnification.
Within the tumor region of ccRCC and pRCC, there are mainly two types of nuclei, endothelial nuclei and tumor nuclei with grade 1 to 3. For each image patch, every nucleus is annotated by three well-trained annotators with OpenHI platform \cite{openHI}. 
After annotation, majority voting is adapted to assign the type of each nucleus, and then these results are reviewed by pathologists. 
This dataset contains 70945 annotated nuclei that consist of 16652 endothelial nuclei and 54293 tumor nuclei (45108, 6406, 2779 for grades 1 to 3, respectively), totally in four classes. 

\subsection{Evaluation Metrics}

%The evaluation of the nuclei grading method needs to be executed from two aspects, the performance of instance segmentation and classification. 
The evaluation of the nuclei grading method needs to be executed in instance segmentation and classification, respectively. 
For instance segmentation, three metrics are applied: the dice coefficient that is a pixel-level metric, the Aggregated Jaccard Index (AJI) that is an object-level metric, and the panoptic quality (PQ) metric that is proposed in \cite{Panoptic}. 
The PQ consists of two parts, detection quality, and segmentation quality. 
%It has been proven to be an accurate quantification and interpretability for nuclei segmentation \cite{Hovernet,gamper2020pannuke}.
It has been proven to be an accurate and interpretable metric for nuclei segmentation \cite{Hovernet,gamper2020pannuke}.

For classification, following the same idea of MoNuSAC \cite{monusac2020}, the PQ of each class (i.e., $PQ_{1}$, $PQ_{2}$, $PQ_{3}$, $PQ_{e}$) and the average PQ (aPQ) are adopted. 
The aPQ is the average of each PQ per class.
% The PQ of each class has the same calculation formula as the overall PQ, but needs to take the segmented instance of one class as the foreground and the other instances as the background.

\subsection{Results}

We randomly divide the ccRCC grading dataset into training (70\%), validation (10\%), and testing (20\%) set, for training and evaluation. 
We implement our method with the open-source library TensorFlow on a work-station with two NVIDIA 2080Ti GPUs. 
For training, data augmentations including flip, rotation, blur are applied to all the models.
The dataset and the source code are available at: \url{https://dataset.chenli.group/home/ccrcc-grading} and \url{https://github.com/ZeyuGaoAi/Composite_High_Resolution_Network}.

\subsubsection{Comparisons with State-of-the-art Methods}

\begin{table}[t]
\caption{Nuclei classification comparison between our method and the state-of-the-art models on ccRCC Grading dataset.}
\label{tab1}
\centering
\begin{tabularx}{\textwidth}{c|*8{>{\centering\arraybackslash}X}}
\hline
Methods   & Dice  & AJI   & PQ   & aPQ   & $PQ_{1}$   & $PQ_{2}$   & $PQ_{3}$   & $PQ_{e}$   \\ \hline
U-Net     & 0.8615 & 0.7344 & 0.6827 & 0.4776 & 0.6001 & 0.3659 & 0.5588 & 0.3857 \\
Mask-RCNN & 0.8657 & 0.7394 & 0.7126 & 0.4749 & 0.6297  & 0.3338 & 0.5423 & 0.3939 \\
Micro-Net & 0.8712 & 0.7375 & 0.7029 & 0.5107 & 0.6432 & 0.3780 & 0.6052 & 0.4163 \\
HoVer-Net & 0.8760 & 0.7440 & 0.7359 & 0.5068 & 0.6319 & 0.3761 & 0.5821 & 0.4370 \\
CHR-Net   & \textbf{0.8790} & \textbf{0.7519} & \textbf{0.7497} & \textbf{0.5458} & \textbf{0.6819} & \textbf{0.4027} & \textbf{0.6271} & \textbf{0.4713} \\ \hline
\end{tabularx}
\end{table}

We compare the proposed CHR-Net with U-Net, Mask-RCNN \cite{maskrcnn}, Micro-Net \cite{micro-net} and HoVer-Net. 
For U-Net and Micro-Net that are originally proposed for segmentation, we use an output dimension of 5 rather than 2. To balance the size of model parameters, the ResNet-34 backbone is used for CHR-Net, and the backbone encoders of U-Net, Mask-RCNN, and HoVer-Net are ResNet-50. 
We initialize the backbone encoders with pre-trained weights on the ImageNet and freeze their parameters for training 50 epochs, then fine-tune all the models for another 50 epochs. 
The adam optimizer with a $10^{-4}$ learning rate ($10^{-5}$ after 25 epochs) is used for the training and fine-tuning. 

The comparison results of each method are shown in Table \ref{tab1}. 
It can be seen that CHR-Net achieves the best performance on both segmentation and classification metrics. 
For nuclei segmentation, both U-Net and Micro-Net show the worse performance because they do not adopt any learning schema for the instance segmentation.  
HoVer-Net has the closest performance to CHR-Net (0.7359 vs. 0.7497 on PQ), but the distance map learned from the original image is sensitive to noise, thus leads to an over-segmentation problem in HoVer-Net.  
For nuclei classification and grading, CHR-Net significantly outperforms any other methods, \textit{i.e.,} U-Net by 6.82\%, Mask-RCNN by 7.09\%, Micro-Net by 3.51\%, and HoVer-Net by 3.9\% on aPQ. 
Note that, Micro-Net achieves the second-best performance among all the methods.
We suppose that the multi-resolution inputs of Micro-Net are beneficial to the ccRCC grading task as same as the auxiliary inputs from our method. 
The qualitative comparisons are provided in the supplementary file.
%For nuclei classification and grading, CHR-Net performs significantly putperform any other methods, \textit{i.e.,} U-Net by 6.82\%, Mask-RCNN by 7.09\%, Micro-Net by 3.51\%, HoVer-Net by 3.9\% on the aPQ. Note that, Micro-Net achieves the second best performance among all the methods, we suppose that the multi-resolution inputs of Micro-Net are benefit to ccRCC grading task, which fits our idea (the auxiliary inputs of different tasks). The qualitative comparisons are provided in the supplementary file.

\subsubsection{Ablation Study}

\begin{table}[t]
\caption{Ablation study of different branches on ccRCC Grading dataset.}
\label{tab2}
\centering
\begin{tabularx}{\textwidth}{c|*8{>{\centering\arraybackslash}X}}
\hline
Branches & Dice           & AJI            & PQ            & aPQ            & $PQ_{1}$   & $PQ_{2}$   & $PQ_{3}$   & $PQ_{e}$            \\ \hline
SHR      & 0.8692          & 0.7233          & 0.6980          & 0.5064         & 0.6376         & 0.3827         & 0.5711         & 0.4342          \\
MHR      & 0.8694          & 0.7244          & 0.7010          & 0.5156         & 0.6324         & \textbf{0.4029}         & 0.5980         & 0.4291          \\
MHR+UDist   & 0.8758       & 0.7280          & 0.7227          & 0.5235          & 0.6601         & 0.3950         & 0.5930         & 0.4460          \\
MHR+WDist  & \textbf{0.8790} & \textbf{0.7519} & \textbf{0.7497} & \textbf{0.5458} & \textbf{0.6819} & 0.4027     & \textbf{0.6271}         & \textbf{0.4713} \\ \hline

\end{tabularx}
\end{table}

%We investigated the three components of the proposed model CHR-Net. Specifically, 
We conduct four ablation experiments by using CHR-Net with different branches. 
SHR indicates the model that only has one HRFE for the four-class nuclei classification without using the segmentation branch. 
MHR represents the model that has two HRFEs for tasks 1 and 2.
UDist and WDist denote two types of segmentation branches which are 1) branch that learns the distance maps from original images with U-Net and 2) branch that predicts the binary maps with W-Net, respectively.  
From Table. \ref{tab2}, we can observe that MHR achieves higher PQ values than SHR on grades 2 and 3, which demonstrates the usefulness of learning the proposed two cross-category classification tasks.  
Due to the additional branches of instance segmentation, MHR+UDist and MHR+WDist have better segmentation performance than MHR. 
Compare with MHR+UDist, MHR+WDist utilizes the two-stage learning strategy to achieve a higher Dice, AJI, and PQ. Meanwhile, affected by the segmentation performance, the PQ values of grade 1, 3, and endothelial from MHR+WDist are also significantly increased, and the average PQ is 3.94\% higher than the base model SHR. 
It is worth mentioning that without using the segmentation branch, the $PQ_{1}$ and $PQ_{e}$ of SHR are much higher than U-Net (3.75\% and 4.85\%), which illustrates the effectiveness of the proposed HRFE.

\section{Conclusion}
In this paper, we proposed a novel nuclei segmentation and classification method, especially for nuclei grading. A corresponding dataset of ccRCC nuclei grading was introduced and evaluated. Integrating with cancer region detection \cite{gao2020renal}, the whole automatic nuclei grading system can reduce the burden of pathologists and further provide quantitative data (see in supplementary file) for cancer research. We hope that the downstream tasks of nuclei segmentation and classification can draw more attention and our work can inspire other researchers with similar challenges.

\subsubsection{Acknowledgements.}
This work has been supported by National Natural Science Foundation of China (61772409); This work has been supported by the National Key Research and Development Program of China (2018YFC0910404); The consulting research project of the Chinese Academy of Engineering (The Online and Offline Mixed Educational Service System for “The Belt and Road” Training in MOOC China); Project of China Knowledge Centre for Engineering Science and Technology; The innovation team from the Ministry of Education (IRT\_17R86); and the Innovative Research Group of the National Natural Science Foundation of China (61721002).
The results shown here are in whole or part based upon data generated by the TCGA Research Network: https://www.cancer.gov/tcga.

\bibliographystyle{splncs04}
\bibliography{refer}

\end{document}